\documentclass{SCGE}
\usepackage[colorlinks,linkcolor=blue,citecolor=blue,urlcolor=blue]{hyperref}

\usepackage[utf8]{inputenc}
\usepackage[T1]{fontenc}
\usepackage{rotating}
\usepackage{footnote}
\newcolumntype{C}[1]{>{\centering\let\newline\\\arraybackslash\hspace{0pt}}m{#1}}
\usepackage[math]{blindtext}
\usepackage{times}
\usepackage{newtxtext,newtxmath}
\usepackage{graphicx}	
\usepackage{amsmath}	
\usepackage{amssymb}	
\usepackage{xspace}
\usepackage{bigints}
\usepackage{siunitx}
\usepackage{diagbox}

\usepackage{amsthm}
\usepackage{mathtools}
\usepackage[normalem]{ulem}
\usepackage{float}
\usepackage{color}
\usepackage[sort&compress,numbers]{natbib}
\usepackage{gensymb}
\usepackage{upgreek}
\usepackage{booktabs}

\usepackage{enumerate}
\usepackage{natbib}
\usepackage{multirow}

\usepackage{enumitem}
\usepackage{hyperref}

\usepackage[utf8]{inputenc}
\DeclareUnicodeCharacter{2212}{\textminus}

\let\citedash\relax
\makeatletter \providecommand{\citedash}{\hbox{-}\penalty\@m}
\makeatother
\begin{document}

\begin{picture}(0,0){\rm
\put(0,-20){\makebox[160truemm][l]{\bf {\sanhao\raisebox{2pt}{.}}
Article  {\sanhao\raisebox{1.5pt}{.}}}}}
\put(0,-34){\jiuwuhao {\textcolor[rgb]{0.5,0.5,0.5}{\sf 
}}}
\end{picture}

\def\bm{\boldsymbol}

\def\dl{\displaystyle}
\def\du{\end{document}}
\def\d{{\rm d}}
\def\e{{\rm e}}
\def\i{{\rm i}}

\Year{2022} %
\Month{XXXX} %
\Vol{xx} %
\No{x} %
\BeginPage{1} %
\AuthorMark{{\rm X. Shan et al.} }  
\DOI{} 
\ArtNo{000000}

\title[An adaptive hierarchical tree algorithm]{Wave effect of gravitational waves intersected with a microlens field II: an adaptive hierarchical tree algorithm and population study}

\author[2,1]{Xikai Shan}{}%
\author[1]{Guoliang Li}{Corresponding author (E-mail: guoliang@pmo.ac.cn)}%
\author[3,4]{Xuechun Chen}{}%
\author[5,6]{Wen Zhao}{}%
\author[3,4]{Bin Hu}{}%
\author[2]{Shude Mao}{}%

\address[{\rm1}]{Purple Mountain Observatory, Chinese Academy of Sciences, Nanjing, Jiangsu, 210023, China;}
\address[{\rm2}]{Department of Astronomy, Tsinghua University, Beijing 100084, China;}
\address[{\rm3}]{Institute for Frontier in Astronomy and Astrophysics, Beijing Normal University, Beijing, 102206, China;}
\address[{\rm4}]{School of Physics and Astronomy, Beijing Normal University, Beijing 100875, China;}
\address[{\rm5}]{School of Astronomy and Space Science, University of Science and Technology of China, Hefei, 230026, China;}
\address[{\rm6}]{CAS Key Laboratory for Researches in Galaxies and Cosmology, Department of Astronomy, University of Science and Technology of China, \\Chinese Academy of Sciences, Hefei, 230026, China}

\maketitle \vspace{-3.5mm}{\footnotesize\begin{center} Received Month date, Year; accepted Month date, Year
\end{center}}\vspace*{-5mm}

\begin{center}
\rule{16.5cm}{0.4pt}
\parbox{16.5cm}
{\begin{abstract}
The gravitational lensing wave effect generated by a microlensing field embedded in a lens galaxy is an inevitable phenomenon in strong lensed gravitational waves (SLGWs). 
This effect presents both challenges and opportunities for the detection and application of SLGWs. 
However, investigating this wave effect requires computing a complete diffraction integral over each microlens in the field.
This is extremely time-consuming due to the large number of microlenses ($10^3\sim10^6$).
Therefore, simply adding all the microlenses is impractical. 
Additionally, the complexity of the time delay surface makes the lens plane resolution a crucial factor in controlling numerical errors.
In this paper, we propose a trapezoid approximation-based adaptive hierarchical tree algorithm to meet the challenges of calculation speed and precision. 
We find that this algorithm accelerates the calculation by four orders of magnitude compared to the simple adding method and is one order of magnitude faster than the fixed hierarchical tree algorithm proposed for electromagnetic microlensing.
More importantly, our algorithm ensures controllable numerical errors, increasing confidence in the results.
Together with our previous work~\citep{Shan:2022xfx}, this paper addresses all numerical issues, including integral convergence, precision, and computational time~\footnote{The C++ source code is publicly available at \url{https://github.com/xkshan97/Microlensing_Wave_Effect}}.
Finally, we conducted a population study on the microlensing wave effect of SLGWs using this algorithm and found that the microlensing wave effect cannot be ignored, especially for Type II SLGWs (from saddle position of the time delay surface) due to their intrinsic geometric structures and their typical intersection with a denser microlensing field.
Statistically, more than $33\%$ ($11\%$) of SLGWs have a mismatch larger than $1\%$ ($3\%$) compared to the unlensed waveform.
Additionally, we found that the mismatch between signal pairs in a doubly imaged GW is generally larger than $10^{-3}$, and $61\%$ ($25\%$) of signal pairs have a mismatch larger than $1\%$ ($3\%$). 
Therefore, the microlensing-induced mismatch can reduce the SLGW identification ability using the overlapping method.
\end{abstract}}
\end{center}\vspace*{-0.6cm}

\begin{center}
\parbox{16.5cm}
{\bf\jiuhao Gravitational lensing, Diffractional integral, Algorithm, Gravitational wave}
\end{center}

\begin{center}
{\PACS{\rm 04.30.−w, 04.30.Tv, 98.62.Sb, 95.85.Sz}}
\end{center}

\textwidth=178truemm \textheight=236truemm

\wuhao\vspace*{1.5mm}

\renewcommand{\baselinestretch}{1.08} \baselineskip 12.2pt\parindent=10.8pt


\section{Introduction}
\label{sec:intro}
Strong lensing gravitational waves (SLGW) as a novel probe have demonstrated numerous powerful cosmological applications~\citep{Li:2018prc, Oguri:2019fix, Hannuksela:2020xor, Cao:2021zpf, Bian:2021ini, Liao:2022gde, Wu:2022vrq, Liu:2023ikc, Wempe:2022zlk}, such as precision Hubble constant measurement~\citep{Liao_2017}, testing modified gravity models~\citep{Collett:2016dey, Andreas:2021zxc, Narola:2023viz}, and studying the nature of dark matter~\citep{Liao:2018ofi, Cao:2022mrc, Tambalo:2022wlm, Seo:2023rjd, Fairbairn:2022xln, Cheung:2024ugg, Zumalacarregui:2024ocb}. 
In contrast to the past decade when GW data were too sparse to detect SLGW events, we are now on the verge of its first event detection, especially for binary neutron star lensing events~\citep{Smith:2022vbp}, which will stimulate multi-messenger lensing applications. 
Therefore, delving deeper into lens modeling and sub-structure effects is valuable to gain a better understanding of GW lensing and its systematics.

The microlensing field, including stars, remnants, and primordial black holes, constitutes one of the most important components of the lens galaxy. 
These objects will introduce a unique lensing wave effect on the SLGW generated by stellar mass binary objects~\citep{2019Diego, Meena:2019ate, 2021Anuj}, owing to the comparable scale of the lens Schwarzschild radius and the wavelength of the GW~\citep{2003Takahashi}. 
Previous works have demonstrated that the microlensing wave effect is a double-edged sword; on one hand, it can bias the parameter estimation of SLGWs and lead to the loss of SLGW identification~\citep{Mishra:2023ddt, Shan:2023qvd}. On the other hand, it can also aid in the identification of SLGWs~\citep{Shan:2023ngi}.
Therefore, conducting more and further studies on the microlensing wave effect will facilitate our understanding of its nature and help find robust methods to convert its disadvantages into advantages. 
However, the first important issue that we need to overcome before application is the diffraction integral calculation problem.

In the past thirty years, pioneers have proposed many algorithms to solve this oscillation integral, including contour integration~\citep{UG95}, asymptotic expansion methods~\citep{press1992numerical}, Levin’s method~\citep{levin1982procedures}, Filon-type methods~\citep{filon1930iii, xiang2007efficient, iserles2006computation}, and two new hybrid algorithms~\citep{guo2020convergence, Tambalo:2022plm, Villarrubia-Rojo:2024xcj}. 
However, directly applying these algorithms to the microlensing field scenario is impractical due to the presence of thousands or even millions of microlenses, leading to inevitable issues with convergence, precision, and computational time. 
In our previous work~\citep{Shan:2022xfx}, we have addressed the convergence problem and improved numerical precision using a component decomposition algorithm; however, issues with lens plane resolution and computational time remain unresolved.

In this paper, we propose an adaptive hierarchical tree algorithm, inspired by the fixed hierarchical tree algorithm designed for the generation of microlensing magnification maps~\citep{Wambsganss1990, zheng2022}. 
In Section~\ref{sec:method}, we will introduce the basic theory of the diffraction integral and our algorithm. 
In Section~\ref{sec:result}, we will test the precision and acceleration of this new algorithm. 
In Section~\ref{sec:population}, we will use our new algorithm to analysis the microlensing wave effect on SLGW from a population perspective.
Finally, in Section~\ref{sec:condis}, we will draw conclusions and discussions.

\section{Method} 
\label{sec:method}
\subsection{Diffraction integral}
\label{subsec:DiffInter}

In the context of the scalar wave approximation, the gravitational wave (GW) lensing effect can be quantified using the diffraction integral~\citep{schneider1992gravitational,10.1143/PTPS.133.137,2003Takahashi} represented by the equation:

\begin{equation}
\label{eq:DiffInter}
F(\omega, \boldsymbol{y})=\frac{2 G \langle \mathrm{M}_L \rangle(1+z_L) \omega}{\pi c^{3} i} \int_{-\infty}^{\infty} d^{2} x \exp \left[i \omega t(\boldsymbol{x}, \boldsymbol{y})\right],
\end{equation}
where $F(\omega, \boldsymbol{y})$ denotes the amplification factor, $\omega$ and $\boldsymbol{y}$ represent the GW's circular frequency and its position (normalized by the Einstein radius) in the source plane, respectively, $\langle \mathrm{M}_L \rangle$ is the average microlens mass, $z_L$ is the lens redshift, and $\boldsymbol{x}$ is the coordinate (normalized by the Einstein radius) in the lens plane.
In Eq.~(\ref{eq:DiffInter}), $t(\boldsymbol{x}, \boldsymbol{y})$ is the time delay function for the microlensing field embedded in the lens galaxy/galaxy cluster scenario, which is given by~\citep{Wambsganss1990, 1992grlebookS, 2021xuechunchen}:
\begin{equation}
\begin{split}
\label{eq:TimeDelay}
t(\boldsymbol{x},\boldsymbol{x}^{i},\boldsymbol{y}=0)&=\underbrace{\frac{k}{2}\left((1-\kappa+\gamma) x_{1}^{2}+(1-\kappa-\gamma) x_{2}^{2}\right)}_{t_\text{smooth}(\kappa,\gamma,\boldsymbol{x})}-\underbrace{\left[\frac{k}{2}\sum_{i}^{N_*} \frac{\mathrm{M}_{L,i}}{\langle \mathrm{M}_L \rangle} \ln \left(\boldsymbol{x}^{i}-\boldsymbol{x}\right)^{2} + k\phi_{-}(\boldsymbol{x})\right]}_{t_\text{micro}(\boldsymbol{x},\boldsymbol{x}^{i})} \ .
\end{split}
\end{equation}
Here, $k=4 G \langle \mathrm{M}_L \rangle(1+z_L)/c^3$, $\mathrm{M}_{L,i}$ and $\boldsymbol{x^{i}}$ are the mass and coordinate of the $i$th microlens, respectively, and $N_*$ is the number of microlenses. 
$\kappa$ and $\gamma$ are the macro lensing convergence and shear.
$t_\text{smooth}(\kappa,\gamma,\boldsymbol{x})$ and $t_\text{micro}(\boldsymbol{x},\boldsymbol{x}^{i})$ are the macro lensing and micro lensing time delay, respectively.
Here, we set the macro image position as the coordinate origin ($\boldsymbol{y} = 0$).
The contribution from a negative mass sheet, denoted by $\phi_{-}(\boldsymbol{x})$, is included to ensure the total convergence $\kappa$ remains unchanged when adding microlenses~\citep{Wambsganss1990, 2021xuechunchen, zheng2022}.

However, due to the oscillatory nature of the integrand, conventional numerical integration methods are often inadequate.
To address this limitation, many works have proposed different algorithms.
For example, \citet{levin1982procedures,press1992numerical,filon1930iii,iserles2006computation,xiang2007efficient,guo2020convergence,Tambalo:2022plm}, etc, proposed effective and fast algorithm for isolated microlens, and \citet{UG95} proposed an general algorithm for both isolated and microlensing field.
In this paper, we will apply~\citet{UG95}'s strategy to solve this oscillation integral.
In detail, this algorithm Fourier transforms the amplification factor into the time domain:
\begin{equation}
\label{eq:DiffTime}
\tilde{F}\left(t, \boldsymbol{y}\right) \equiv \frac{1}{2 \pi} \int_{-\infty}^{\infty} \mathrm{d} \omega \exp \left(-i \omega t\right) \frac{F(\omega, \boldsymbol{y})}{C_{\omega}} \ ,
\end{equation}
where $\tilde{F}\left(t, \boldsymbol{y}\right)$ is the time domain amplification factor and $C_{\omega}$ is the coefficient before the integrand of Eq.~(\ref{eq:DiffInter}):
\begin{equation}
C_{\omega}=\frac{2 G \langle \mathrm{M}_L \rangle(1+z_L) \omega}{\pi c^{3} i} \ .
\end{equation}
Upon substituting Eq.~(\ref{eq:DiffInter}) into Eq.~(\ref{eq:DiffTime}), it is evident that the time domain amplification factor is proportional to the time delay probability density function in the lens plane:
\begin{equation}
\label{eq:TimeDomainMag}
\tilde{F}\left(t, \boldsymbol{y}\right) =\int_{-\infty}^{\infty} \mathrm{d}^{2} x \delta\left[t(\boldsymbol{x}, \boldsymbol{y})-t\right]=\frac{|\mathrm{d} S|}{\mathrm{d} t } \ ,
\end{equation}
which can be determined by calculating the surface area $\mathrm{d} S$ within the time delay interval $\mathrm{d} t $.
The subsequent step is to inversely Fourier transform $\tilde{F}\left(t, \boldsymbol{y}\right)$ to recover $F(\omega, \boldsymbol{y})$.
One can find that this algorithm can efficiently eliminates the oscillation problem.

To calculate the time domain amplification, one direct and straightforward method is to pixelate the lens plane and then accumulate the area of the pixel within the same time delay interval $[t - \Delta t/2, t + \Delta t /2]$~\citep{2019Diego}. 
This method offers an advantage compared to the contour integration method~\citep{UG95, 2021Anuj}. 
Specifically, this method eliminates the need to identify the image position, where solving a nonlinear lens function is required, which can be time-consuming in the context of the microlensing field scenario.
However, this Simple Adding (refer to as SA in the later of this paper) method needs an extremely high-resolution pixelized lens plane to avoid numerical error induced by the time delay variation in the inner region of the pixel.
In other words, if the maximum time delay difference in the pixel is larger than the time delay resolution, simply arranging the whole pixel area into one time delay interval will introduce numerical error.
Therefore, in the case of a highly dense field configuration, the total calculation time, 
\begin{equation}
\label{eq:tsa}
t_\mathrm{SA} = N_\mathrm{pixel} \cdot N_*,
\end{equation}
can be long.
Here, $N_\mathrm{pixel}$ is the total number of pixels in the lens plane and $N_*$ is the number of microlenses.

To reduce the numerical errors induced by the pixels that cross the time delay bins, and to speed up the calculation by avoiding an extremely high resolution pixelized lens plane, we propose a trapezoid approximation-based adaptive hierarchical tree algorithm.
Hereafter, we will refer to this algorithm as TAAH (\textbf{T}rapezoid \textbf{A}pproximation-based \textbf{A}daptive \textbf{H}ierarchical) tree algorithm.

\subsection{Trapezoid approximation in pixel}
\begin{figure}
\centering
\includegraphics[width=0.4\columnwidth]{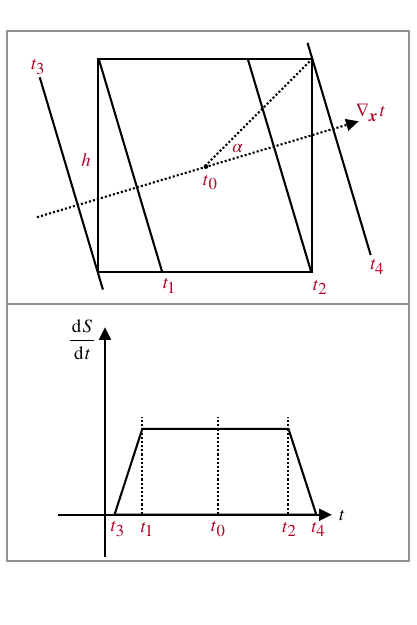}
\caption{This figure illustrates the time delay gradient (upper panel) and the time domain amplification factor (lower panel) within a sufficiently small pixel that does not include any microlenses. 
$h$ is the side length of the pixel.
$t_0$ represents the time delay at the middle of this pixel, and the straight lines labeled $t_{1\sim4}$ are the iso-time delay curves. 
The dashed straight line with an arrow depicts the time delay gradient within this pixel, and $\alpha$ is the angle between the gradient and the diagonal.}
\label{fig:Trap}
\end{figure}
\label{subsec:Trape}
The time domain amplification factor within a pixel can be evaluated through the contour integral:

\begin{equation}
\frac{|\mathrm{d} S|}{\mathrm{d} t} = \oint_C \frac{\mathrm{d} s}{\left|\nabla_{\boldsymbol{x}} t(\boldsymbol{x}, \boldsymbol{y}) \right|} \ .
\end{equation}
Here, $\mathrm{d} s$ represents the line element, and $C$ is the length of the iso-time delay curve. 
The gradient of the time delay can be expressed as:
\begin{equation}
\label{eq:tgrad}
\begin{split}
\nabla_{\boldsymbol{x}} t(\boldsymbol{x}, \boldsymbol{y})&=k\left(\begin{array}{cc}1-\kappa+\gamma & 0 \\ 0 & 1-\kappa-\gamma\end{array}\right) \boldsymbol{x}-k\sum_{i}^{N_{*}} \frac{\mathrm{M}_{L,i}}{\langle \mathrm{M}_L \rangle} \frac{\boldsymbol{x}-\boldsymbol{x}^{(i)}}{\left|\boldsymbol{x}-\boldsymbol{x}^{(i)}\right|^{2}} - k\nabla_{\boldsymbol{x}}\phi_{-}(\boldsymbol{x}) \ ,
\end{split}
\end{equation}
and more details can be found in~\citet{zheng2022}. 
It is worth noting that Eq.~(\ref{eq:tgrad}) is differentiable except for the positions of the microlenses.
Therefore, the variation of the gradient of the time delay, denoted as $|\Delta \nabla_{\boldsymbol{x}} t(\boldsymbol{x}, \boldsymbol{y})|$, is finite if the pixel is sufficiently small and does not include any microlens. 
Consequently, the gradient of the time delay can be treated as constant in a sufficiently small pixel. 
Based on this property, we present a sketch that describes the time delay gradient and time domain amplification factor within this type of pixel in Fig.~\ref{fig:Trap}.

In the upper panel, we use a black dashed curve with an arrow to represent the gradient of the time delay and use the black solid curves orthogonal to the dashed curve, labeled with different red tags to represent the iso-time delay curves.
$h$ is the side length of the pixel.
$t_0$ is the time delay in the middle of the square pixel, and $\alpha$ is the angle between the direction of the gradient and the diagonal of the pixel. 
One can find that the contour length between $t_1$ and $t_2$ is constant, and the contour length between $t_1$ and $t_3$ and between $t_2$ and $t_4$ are monotonically decreasing.
Therefore, the time domain amplification factor within the pixel can be described by:
\begin{equation}
\begin{array}{l}
\label{eq:Trape}
\frac{|\mathrm{d} s|}{\mathrm{d} t'}=  \left\{\begin{array}{ll} 0, & t' < t_3 \\ \text{const} \times \frac{t' - t_3}{t_1 - t_3}, & t_3 \leq t' \leq t_1 \\ {\text{const}}, & t_1 < t'< t_2 \\ \text{const} \times \frac{t - t_2}{t_4 - t_2}, & t_2 \leq t' \leq t_4 \\ 0, & t_4 < t' \end{array}\right.\end{array} \\,
\end{equation}
where $\text{const} = \frac{h}{\cos(\pi/4 - \alpha) \times |\nabla_{\boldsymbol{x}} t|}$. 
The lower panel illustrates this theoretical distribution of the time domain amplification factor within this pixel.

For pixels that include microlenses, it can be observed that the span of the time delay in such pixels will be infinite. Consequently, the time domain amplification factor, Eq.~(\ref{eq:TimeDomainMag}), within this type of pixel will be zero. 
Therefore, the trapezoid approximation error in such pixels can be eliminated by reducing the size of the pixel.

After calculating all the time domain amplification factors in all pixels, the final time domain amplification factor at time $t$ in the entire lens plane can be obtained using the formula:
\begin{equation}
\label{eq:td_amp}
\frac{|\mathrm{d} S|}{\mathrm{d} t} |_{t} = \frac{\sum_{i = 0}^{N_\mathrm{pixel}} \int_{t - \Delta t / 2}^{t + \Delta t / 2} \frac{|\mathrm{d} s|}{\mathrm{d} t'}|_{i} \mathrm{d} t'}{\Delta t} \ ,
\end{equation}
where $i$ represents the $i$th pixel, and $N_\mathrm{pixel}$ is the total number of pixels in the image plane.

\subsection{An adaptive hierarchical tree algorithm for diffraction integral}
\label{subsec:adapt}
\begin{figure*}
\centering
\includegraphics[width=0.8\textwidth]{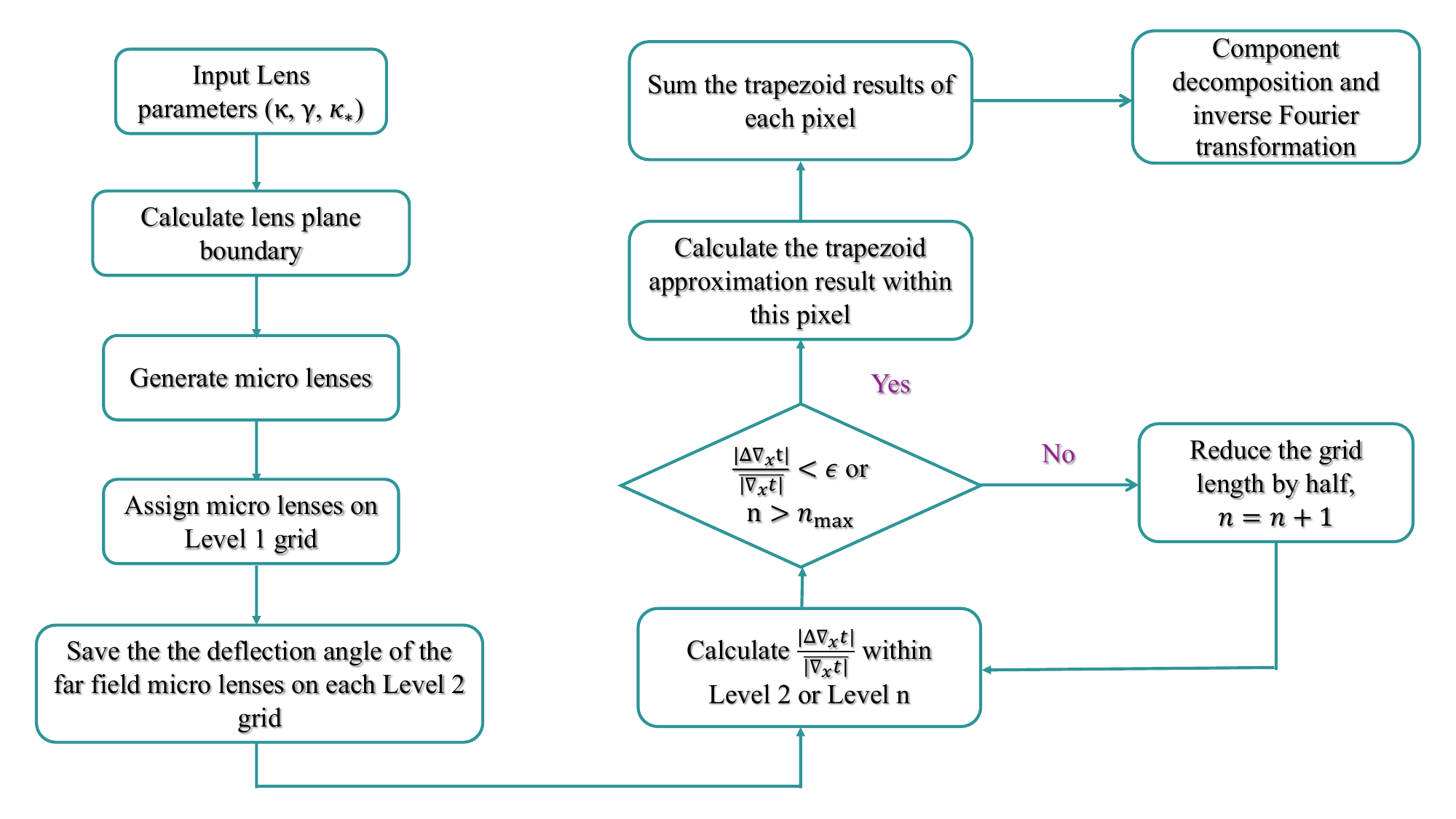}

\caption{Flow chart of the Trapezoid Approximation-based Adaptive Hierarchical (TAAH) tree algorithm.
        }
\label{fig:FlowChart}
\end{figure*}
In Section~\ref{subsec:DiffInter}, we know that the calculation time $t_\mathrm{SA}$ for the microlensing field diffraction integral is proportional to $N_\mathrm{pixel} \times N_*$.
This number is particularly unfavorable for the high-density stellar mass microlensing field.
Therefore, it is impossible to simply sum over all of the microlenses' contribution for each pixel.
In this section, we introduce how our TAAH tree algorithm reduces time consumption.

The TAAH tree algorithm include two fixed grid levels (Level $1$ and Level $2$) and an adaptive grid (Level $n$, $n\geq2$).
The first two fixed grid is based on the algorithm proposed in~\citet{zheng2022}, which is focused on generating microlensing magnification maps.
In detail, the Level $1$ grid stores the indices of the microlenses.
This grid is used to construct a lookup table for quick microlens searching.

Here, we set the resolution of Level $1$ as $L_1 = \min \{L_0/10, L_0/\sqrt{N_*}\}$, where $L_0$ is the length of the total lens plane. 
$L_0/\sqrt{N_*}$ is the average length occupied by each microlens. 
This setting ensures that the number of Level $1$ pixels is comparable to the number of microlenses when the microlens number is relatively large. 
When the number of microlenses is relatively small, we set the length of the Level $1$ grid as $L_0/10$. 
The reason for this choice is that the Level $1$ grid is also used to determine whether the microlens is in the far field or near field. 
This setting ensures the size of the Level $1$ grid is not too big to be used as a standard for far/near field microlens decision.

The Level $2$ grid stores the deflection angle of the far field microlenses. 
As outlined in~\citet{Wambsganss1990, 2021xuechunchen, zheng2022}, the evolution of the lens potential induced by far field microlenses within the Level $2$ grid is smooth, and the distance between these microlenses and the center of the Level $2$ grid is significantly larger than the resolution of the Level $2$ grid. 
Consequently, the time delay and deflection angle of far field microlenses within the Level $2$ grid (in other words, Level $n$ grid in TAAH tree algorithm) can be effectively interpolated using the stored deflection angles (see~\citealt{Wambsganss1990} for more details).
Hence, for a more accurate Level $n$ grid within Level $2$, there is no need to recompute the far-field microlenses, and we only need to use the interpolation polynomial to recover its potential. 
Then, one only needs to sum over the near-field microlenses and add them to the interpolation result to obtain the final result. 
This leads to a substantial reduction in the total computing time:
\begin{equation}
\label{eq:tfixed}
t_\mathrm{FH} \simeq N_{L_2} \cdot N_* + N_\mathrm{pixel} \cdot N_\mathrm{*, near}  \\, 
\end{equation}
where $N_{L_2} \ll N_\mathrm{pixel}$ and $N_\mathrm{*, near} \ll N_*$ in a typical microlensing system.
Here, we use the same resolution of the Level $2$ grid as in~\citet{zheng2022}, which is $L_2 = L_1 / 20$ and far-field microlenses are the lenses outside the nearest eight $L_1$ grids surrounding the $L_1$ grid in which the $L_2$ grid resides (so, the number of near field microlenses $N_\mathrm{*,near}\simeq 9$).
Eq.~(\ref{eq:tfixed}) is the calculation time for the Fixed Hierarchical (FH) tree algorithm introduced in~\citet{zheng2022} for the generation of the microlensing magnification map.

The refined grid relative to Level $2$ is an adaptive Level $n$ grid, which differs from the Level $3$ grid used in~\citet{zheng2022}. 
This choice is inspired by the property of the trapezoid approximation. 
As established in Section~\ref{subsec:Trape}, the time domain amplification factor within the pixel can be approximated by a trapezoid-shaped distribution. 
The accuracy of this approximation relies on the relative changing rate of the gradient of the time delay within the pixel. 
Therefore, if this relative changing rate $|\Delta \nabla_{\boldsymbol{x}} t| / \overline{|\nabla_{\boldsymbol{x}} t|}$ is less than a tiny value $\epsilon$, the approximation is accurate within the tolerance of the error.

Based on this property, we set the length of the level $n$ grid by requiring $|\Delta \nabla_{\boldsymbol{x}} t| / \overline{|\nabla_{\boldsymbol{x}} t|} < \epsilon$. 
In detail, first, we calculate $e_2 = |\Delta \nabla_{\boldsymbol{x}} t| / \overline{|\nabla_{\boldsymbol{x}} t|}$ in the Level $2$ grid. 
If $e_2 < \epsilon$, the trapezoid approximation in the Level $2$ grid is accurate enough, and it does not need further refinement; otherwise, we reduce the grid length by half and calculate $e_n = |\Delta \nabla_{\boldsymbol{x}} t| / \overline{|\nabla_{\boldsymbol{x}} t|}$ in each subpixel (where $n>2$ is the resolution level number, and the number of $e_n$ is $4^{n-2}$). 
The refinement stops when all $e_n$ are less than $\epsilon$.
In each pixel, we use the maximum difference of \( \nabla_{\boldsymbol{x}} t \) among five positions (four vertex positions and the center position) to approximate \( |\Delta \nabla_{\boldsymbol{x}} t| \) , and the mean value of \( \nabla_{\boldsymbol{x}} t \) at these five positions to represent \( \overline{|\nabla_{\boldsymbol{x}} t|} \).

In this process, there are three kinds of special positions that will encounter infinite refinement. 
The first one is our previously mentioned microlens positions, the second one is the origin of the lens plane, and the third one is the microlensing image positions. 
In these three cases, the relative difference within the pixel will be $\infty$ and will not converge by reducing the length of the pixel. 
To avoid infinite refinement, we set a maximum resolution level $n_\mathrm{max}$ to break off the refinement. 
In the results section, one find that the numerical error caused by the trapezoid approximation in these three kinds of pixels is tiny.
These steps complete the construction of the TAAH tree algorithm.

In Fig.~\ref{fig:FlowChart}, we present the flow chart of this algorithm, encompassing the following nine main steps:
\begin{itemize}[label=\textbullet, font=\Large]
\item Input lens parameters, including macro lensing $\kappa$, $\gamma$, and micro lensing $\kappa_*$.
\item Use the method introduced in~\citet{Shan:2022xfx} to calculate the microlensing field boundary $L_0/2$.
\item Randomly place $N_*=\kappa_{*} \times L_0^2 / \pi$ microlenses within the microlensing field boundary box.
\item Set Level $1$ grid with a length $L_1 = \min\{L_0/10, L_0/\sqrt{N_*}\}$, and assign each microlens to a Level $1$ grid.
\item Set Level $2$ grid with a length $L_2 = L_1 / 20$, and store the deflection angle contributed by the far field microlenses.
\item Calculate $|\Delta \nabla_{\boldsymbol{x}} t| / \overline{|\nabla_{\boldsymbol{x}} t|}$ within the Level $2$ or Level $n$ grid. If $|\Delta \nabla_{\boldsymbol{x}} t| / \overline{|\nabla_{\boldsymbol{x}} t|} < \epsilon$ or $n > n_\mathrm{max}$, proceed to the next step; otherwise, reduce the grid length by half and $n = n + 1$.
\item Calculate the trapezoid approximation result within the Level $n$ pixel using Eq.~(\ref{eq:Trape}).
\item Sum over the trapezoid approximation results at all Level $n$ grids using Eq.~(\ref{eq:td_amp}).
\item Use the Component Decomposition algorithm proposed in~\citet{Shan:2022xfx} to recover the frequency domain amplification factor $F(\omega)$.
\end{itemize}

In the final part of this section, we want to highlight the acceleration of this adaptive algorithm.
Compared with the FH tree algorithm~\citep{2021xuechunchen, zheng2022} introduced for the generation of the microlensing magnification map, our algorithm does not require a uniformly refined grid over Level $2$. Therefore, our algorithm can further accelerate the calculation by reducing the second term of Eq.~(\ref{eq:tfixed}),
\begin{equation}
\label{eq:tTAAH}
t_\mathrm{TAAH} \simeq N_{L_2} \times N_* + N_{L_n} \times N_\mathrm{*, near} \ ,
\end{equation}
where $N_{L_n}$ is the total pixel number of the adaptive grid. 
One can find that if $N_{L_n} \ll N_\mathrm{pixel}$, the acceleration will be significant.

\section{Algorithm test}
\label{sec:result}

\subsection{Precision test}
In this section, we evaluate the accuracy of our TAAH tree algorithm by comparing its results with analytical solutions.
We consider three distinct macro lensing scenarios, and the corresponding macro and micro lensing parameters are provided in Table~\ref{ta:OneMicro}.

\begin{table}
  \centering
  \caption{\label{ta:OneMicro} This table listed the macro and micro parameters for algorithm precision test.
  $\kappa$ and $\gamma$ are macro lensing convergence and shear, Micro $x_1$ and Micro $x_2$ are microlens coordinate.
  Here, we set the mass of microlens as $100~\mathrm{M}_\odot$ for Type I (minimum) and Type III (maximum) scenario.}
   \begin{tabular}{c|cccc}
    \hline
    \hline
     & $\kappa$ & $\gamma$ & Micro $x_1$ & Micro $x_2$ \\ 
    \hline
    Type I & $0.7$ & $0$ & $0.1$ & $0$  \\
    Type II & $0.875$ & $0.325$ & none & none \\
    Type III & $1.3$ & $0$ & $10$ & 0 \\
    \hline
    \hline
  \end{tabular}
\end{table}

Specifically, for Type I (minimum) and Type III (maximum) macro lensing images, we set the shear ($\gamma$) value to $0$. 
This choice is made because when a micro lens is embedded in external convergence without external shear, the diffraction has an analytical solution described by~\citep{Shan:2022xfx}:
\begin{equation}
\label{eq:embededin}
\begin{split} 
F(w,\boldsymbol{x})=&-\frac{1}{\lambda^2 w} 2^{-2-\frac{i w}{2}}|w|^{-1+\frac{i w}{2}} \Gamma\left(1-\frac{i w}{2}\right)|\lambda|^{-1+\frac{i w}{2}} e^{\frac{i w}{2}\left(\phi_{m} + \lambda(x_1^2 + x_2^2)\right)} \quad\left((w+2 i)\left|\lambda\right|^{3}|w|^{3}\left(x_{1}^{2}+x_{2}^{2}\right)\right.
\\ &\left(\sinh \left(\frac{\pi w}{4}\right) \operatorname{sgn}(\lambda w)+\cosh \left(\frac{\pi w}{4}\right)\right){ }_{2} F_{3}\left(1-\frac{i w}{4}, \frac{3}{2}-\frac{i w}{4} ; 1, \frac{3}{2}, \frac{3}{2} ;-\frac{1}{16}\lambda^2 w^{2}\left(x_{1}^{2}+x_{2}^{2}\right)^{2}\right) 
\\ &\left.-4 \lambda^2w^2\left(\cosh \left(\frac{\pi w}{4}\right) \operatorname{sgn}(\lambda w)+\sinh \left(\frac{\pi w}{4}\right)\right){ }_{2} F_{3}\left(\frac{1}{2}-\frac{i w}{4}, 1-\frac{i w}{4} ; \frac{1}{2}, \frac{1}{2}, 1 ;-\frac{1}{16}\lambda^2 w^{2}\left(x_{1}^{2}+x_{2}^{2}\right)^{2}\right)\right) \ ,
\end{split}
\end{equation}
where $\lambda = 1 - \kappa$, $w = k \omega$ is the dimensionless frequency, $\Gamma$ and ${}_2F_3$ are gamma function and generalized hypergeometric function.

For Type II (saddle) macro-lensing images, we exclusively test our algorithm under a pure macro-lensing scenario.
This choice is deliberate because logarithmic divergence is the most important characteristic of Type II, and there are no analytical solutions when any micro lens is embedded.

\begin{figure*}
\centering
\includegraphics[width=0.8\textwidth]{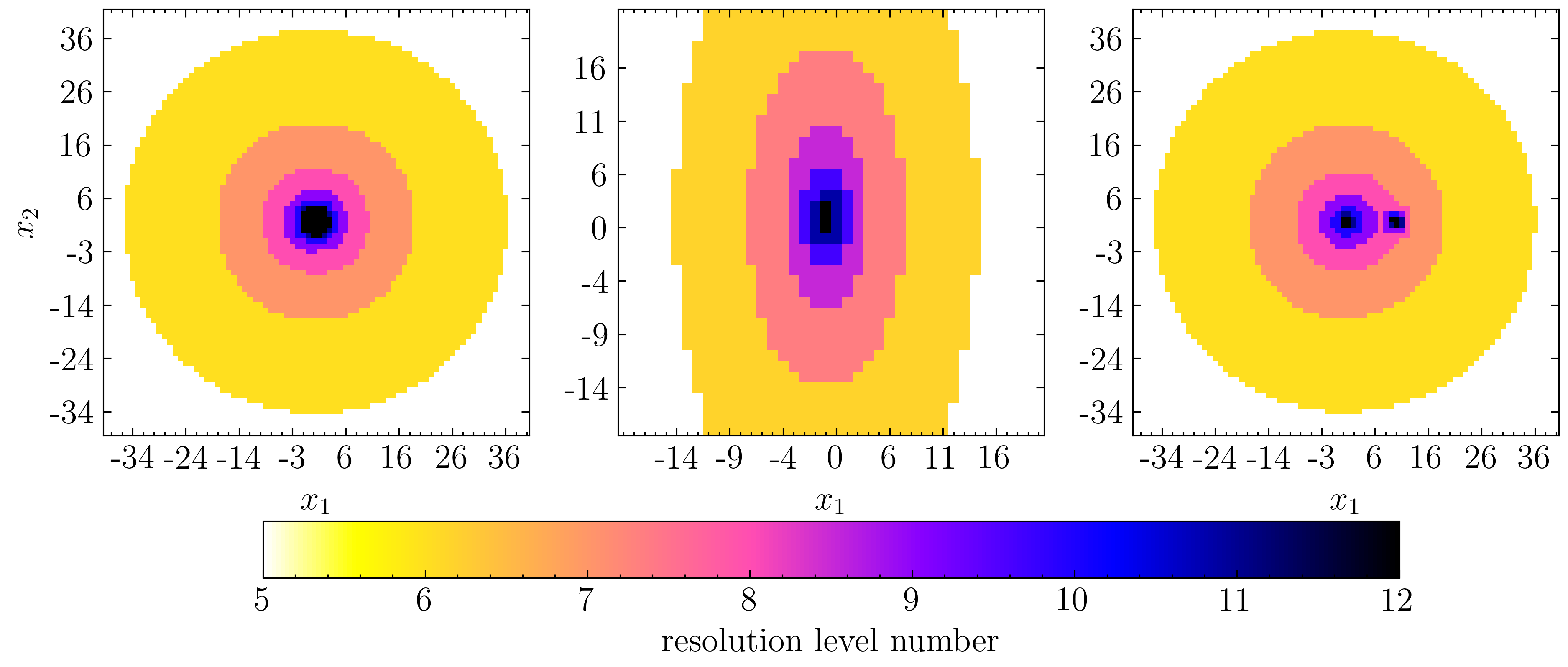}

\caption{Resolution level number $n$ in the lens plane. Columns from left to right are Type I (minimum), Type II (saddle) and Type III (maximum) scenarios. 
The parameters used in these simulations are listed in Table~\ref{ta:OneMicro}. 
}
\label{fig:OneMicroMesh}
\end{figure*}

Fig.~\ref{fig:OneMicroMesh} illustrates the resolution level number $n$ (adaptive Level $n$ grid) in the lens plane for three different scenarios, with Type I on the left, Type II in the middle, and Type III on the right. 
Here, we set $L_2 = 1.024$, $\epsilon = 0.01$, and $n_\mathrm{max} = 12$, resulting in a minimum pixel length of $L_{12} = 0.001$. 
In this figure, the inner region exhibits a higher resolution number, primarily due to macro lensing effects, as $|\Delta \nabla_{\boldsymbol{x}} t| / \overline{|\nabla_{\boldsymbol{x}} t|} \propto 1/|\boldsymbol{x}|$, where $|\boldsymbol{x}|$ is the distance from the center. 
Additionally, the third panel shows that at the micro lens positions, the resolution level is very high, as this region is dominated by micro lensing effects and $|\Delta \nabla_{\boldsymbol{x}} t| / \overline{|\nabla_{\boldsymbol{x}} t|} \propto 1/|\boldsymbol{x} - \boldsymbol{x_\mathrm{micro}}|$, where $\boldsymbol{x_\mathrm{micro}}$ is the coordinate of the micro lenses.
This refinement results are aligned with our discussions in Section~\ref{subsec:adapt}.

\begin{figure*}
\centering
\includegraphics[width=0.8\textwidth]{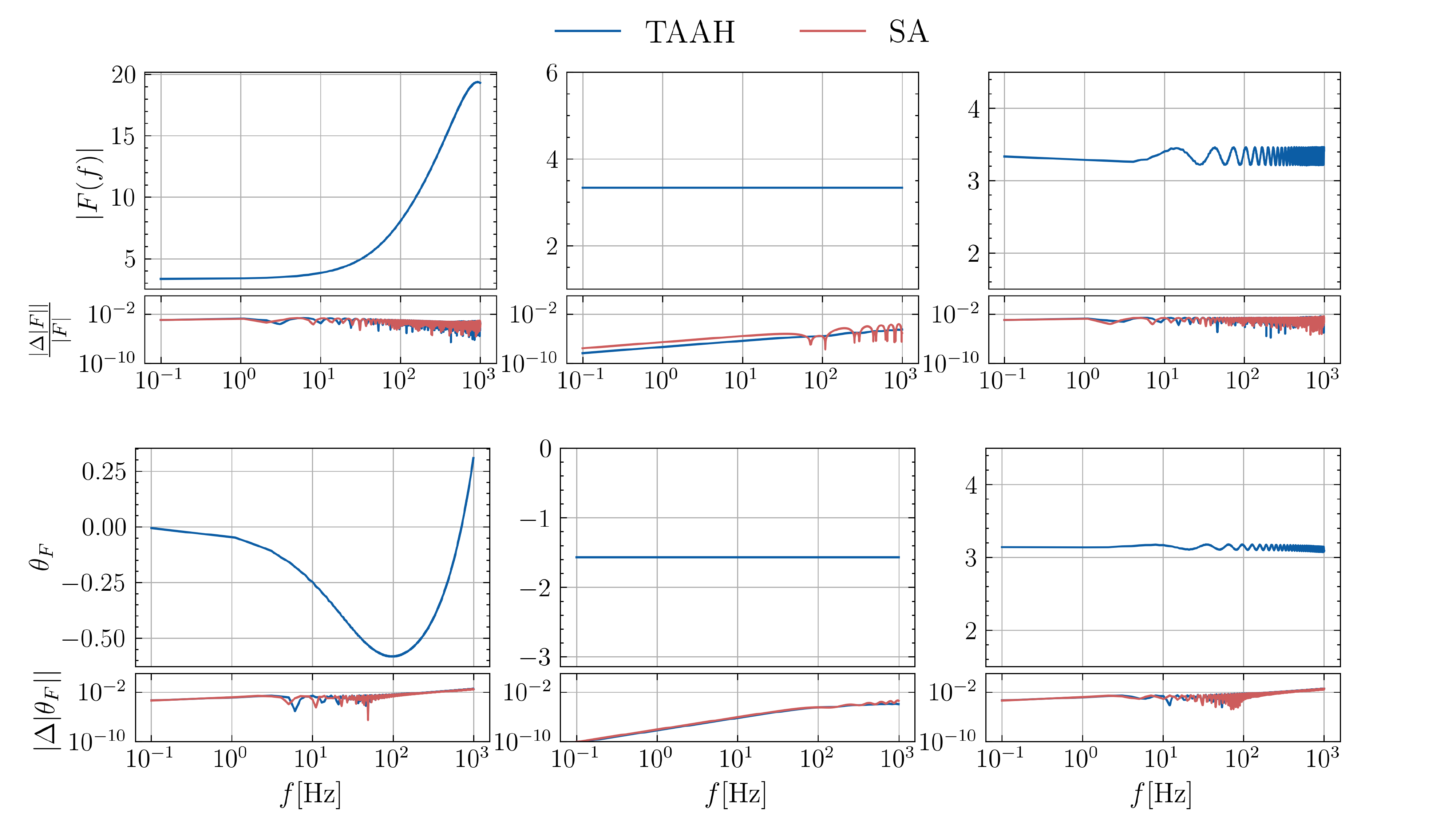}

\caption{Diffraction integral results for three macro + micro or pure macro lensing scenarios. 
Each column corresponds to a different macro lensing image, with Type I (minimum) on the left, Type II (saddle) in the middle, and Type III (maximum) on the right. 
The top (bottom) two rows display the absolute value (phase) of the amplification factor and its relative error between numerical and analytical results.
The blue curves represent our Trapezoid Approximation-based Adaptive Hierarchical (TAAH) tree algorithm, while the red curves represent the Simple Adding (SA) algorithm.
The parameters used in these simulations are listed in Table~\ref{ta:OneMicro}. }
\label{fig:OneMicroFf_ThetaF}
\end{figure*}

Fig.~\ref{fig:OneMicroFf_ThetaF} illustrates the results of the diffraction integral for the three macro + micro/pure macro lensing configurations. 
Each column corresponds to a different macro lensing image, with Type I on the left, Type II in the middle, and Type III on the right. 
The top (bottom) two rows display the absolute value (phase) of the amplification factor and its relative error between numerical and analytical results.
The blue curves represent our TAAH tree algorithm, while the red curves represent the SA (Simple Adding) algorithm. 
For the SA algorithm, the resolution at the lens plane was set to $0.001$, equivalent to the grid length of the maximum resolution level of the TAAH tree algorithm. 
It is evident that the precision of our method is comparable to or more accurate than the SA method but requires less time due to the use of fewer pixels in the lens plane. 
Therefore, the trapezoid approximation can effectively reduce the numerical error arising from the time delay variation within the pixel.
Here, we do not emphasize the algorithm acceleration, as this is a simple case primarily intended to test our algorithm's accuracy.

\subsection{Acceleration test}
In this section, we test the acceleration of the TAAH tree algorithm using realistic microlensing field configurations.
Here, we also consider three different macro lensing image types: Type I, Type II, and Type III. 
The parameters used in the calculations are listed in Table~\ref{ta:MultiPara}. 
The lens plane boundary $L_0$ is determined using the method introduced in~\citet{Shan:2022xfx}.
We choose a precision parameter $\epsilon=0.1$ and set the maximum grid level $n_\mathrm{max} = 9$ (resulting in a resolution of $L_9 = L_2/128$).

\begin{table}
  \centering
  \caption{\label{ta:MultiPara} This table lists the macro and micro lensing field parameters used in simulations.
  $\kappa$ and $\gamma$ are the macro convergence and shear. $\kappa_{*}$ is the convergence of stellar microlensing field. 
  $L_0$ is the lens plane boundary. 
  $\epsilon$ is the precision parameter. 
  Here, we set the mass of all microlenses to $1~\mathrm{M}\odot$.
  }
   \begin{tabular}{c|ccccc} 
    \hline
    \hline
    Parameter & $\kappa$ & $\gamma$ & $\kappa_{*}$ & $L_0$ & $\epsilon$ \\ 
    \hline
    Type I & $0.7$  & $-0.25$ & $0.06$ & $333.3$ & $0.1$\\
    Type II & $0.8$  & $0.25$ & $0.06$ & $433$ & $0.1$\\
    Type III & $1.2$  & $-0.15$ & $0.06$ & $333.3$ & $0.1$  \\
    \hline
    \hline
  \end{tabular}
\end{table}

\begin{figure*}
\centering
\includegraphics[width=0.8\textwidth]{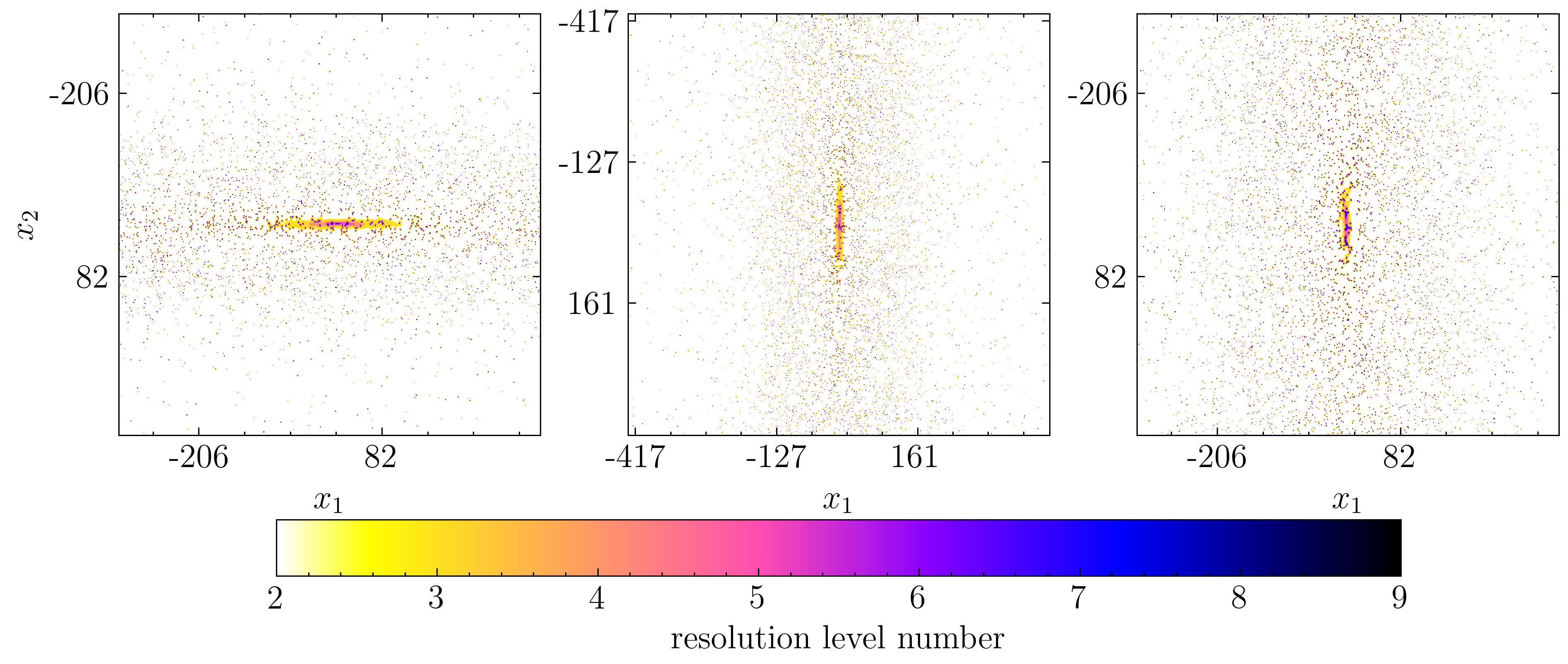}

\caption{Resolution level numbers for microlensing field scenarios. 
Columns from left to right represent Type I (minimum), Type II (saddle), and Type III (maximum) macro images.
The parameters used in these simulations are listed in Table~\ref{ta:MultiPara}.}
\label{fig:MicroFieldMesh}
\end{figure*}

In Fig.~\ref{fig:MicroFieldMesh}, we illustrate the resolution levels for these three macro + micro lensing field configurations. 
It is evident that the majority of grids are at Level $2$ resolution (depicted in black), with only a small portion of the region requiring higher resolution. 
As a result, our TAAH tree algorithm significantly enhances computational efficiency compared to the FH tree algorithm's uniformly refined strategy.

In Table~\ref{ta:MultiResult}, we provide a comparison of calculation times between the SA, FH tree, and TAAH tree algorithms.
$N_{L_2}$ denotes the pixel number of the Level $2$ grid, $N_{L_n}$ represents the total pixel number ($n>2$) for the TAAH tree algorithm, and $N_\mathrm{pixel}$ is the pixel number for the SA and FH tree algorithms, as defined in Eq.~(\ref{eq:tsa}) and Eq.~(\ref{eq:tfixed}).
Here, we set the resolution in the SA and FH tree algorithms equivalent to the minimum resolution in the TAAH tree algorithm, i.e., $L_\mathrm{pixel} = L_9$.
$N_*$ is the number of microlenses. 
The sixth column represents the calculation speedup of the TAAH tree algorithm relative to the SA method, and the seventh column represents the calculation speedup of the TAAH tree algorithm relative to the FH tree algorithm.

From this table, it is evident that the TAAH tree algorithm demonstrates a speedup of approximately four orders of magnitude compared to the SA method and approximately one order of magnitude compared to the FH tree algorithm.

\begin{figure*}
\centering
\includegraphics[width=0.8\textwidth]{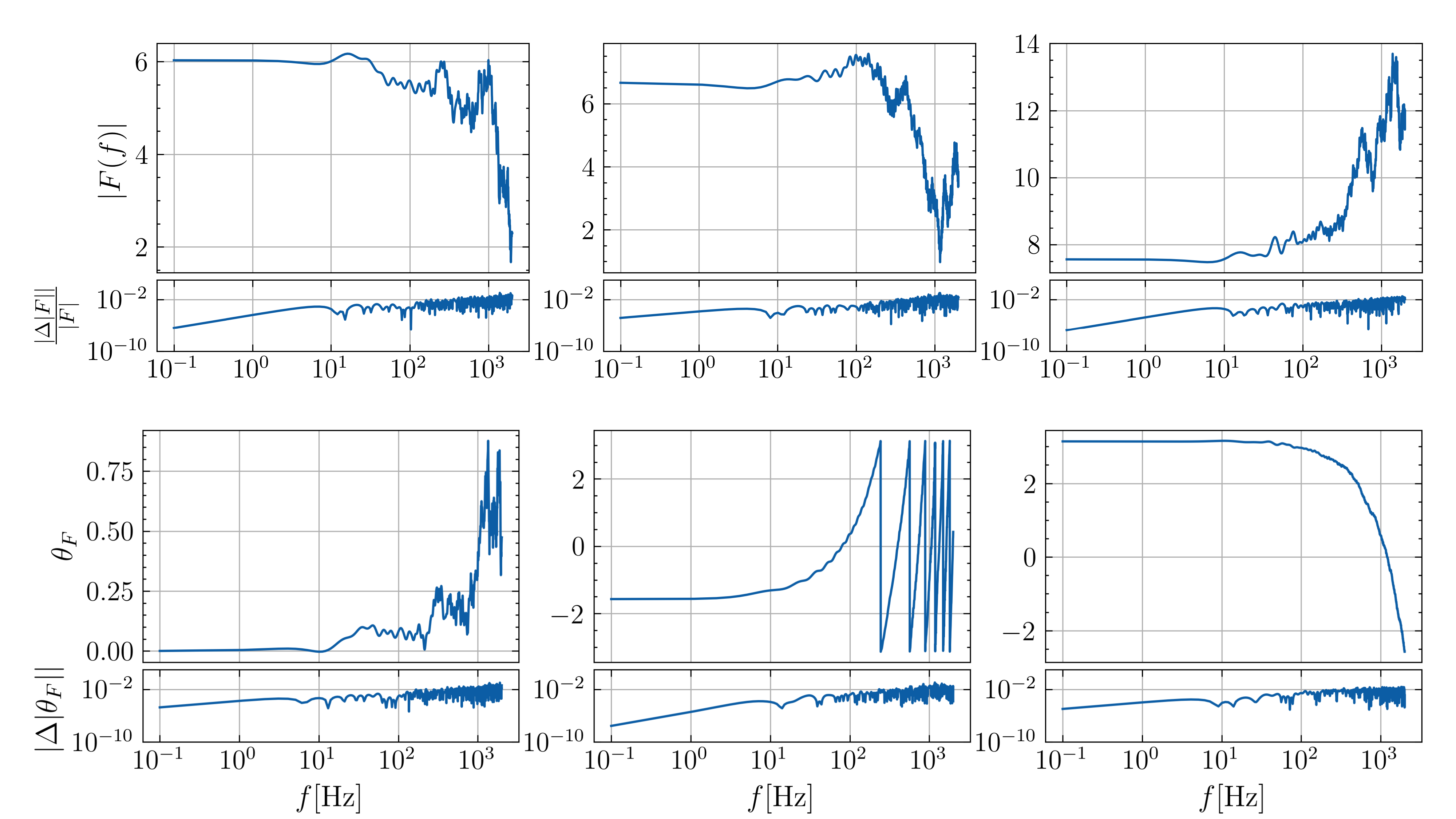}

\caption{Diffraction integral results for the microlensing field. Columns from left to right represent Type I (minimum), Type II (saddle), and Type III (maximum) macro lensing scenarios. The top (bottom) two rows are the absolute (phase) value of the amplification factor using Trapezoid Approximation-based Adaptive Hierarchical (TAAH) tree algorithm and the differences between TAAH tree algorithm and the Fixed Hierarchical (FH) tree algorithm.
The lens parameters used in these simulations are listed in Table~\ref{ta:MultiPara}.}
\label{fig:MicroFieldFf_ThetaF}
\end{figure*}

In Fig.~\ref{fig:MicroFieldFf_ThetaF}, we compare the diffraction integral results of our TAAH tree algorithm with those of the FH tree algorithm. 
We have not presented the results obtained using the SA method due to its excessively long calculation time, which makes it impractical to complete within an acceptable timeframe. 
In this figure, different columns represent different macro lensing image types: Type I on the left, Type II in the middle, and Type III on the right. 
The top two rows depict the absolute values of the amplification factor, while the bottom two rows show the phase values, both using the TAAH tree algorithm and the differences between the TAAH tree algorithm and the FH tree algorithm. 
It can be observed that the differences between these two algorithms are less than $1\%$. 
Therefore, implementing adaptive refinement over the Level $2$ grid based on the trapezoid approximation proves to be efficient, achieving both high precision and time savings.

\begin{table*}
  \centering
  \caption{\label{ta:MultiResult} 
  This table lists the grid number and calculation time comparisons between the Simple Adding (SA), Fixed Hierarchical (FH) tree, and Trapezoid Approximation-based Adaptive Hierarchical (TAAH) tree algorithms.
$N_{L_2}$ denotes the pixel number of the Level $2$ grid, $N_{L_n}$ represents the total pixel number ($n>2$) for the TAAH tree algorithm, and $N_\mathrm{pixel}$ is the pixel number for the SA and FH tree algorithms, as defined in Eq.~(\ref{eq:tsa}) and Eq.~(\ref{eq:tfixed}).
Here, we set the resolution in the SA and FH tree algorithms equivalent to the minimum resolution in the TAAH tree algorithm, i.e., $L_\mathrm{pixel} = L_9$.
$N_*$ is the number of microlenses. 
The sixth column represents the calculation speedup of the TAAH tree algorithm relative to the SA method, and the seventh column represents the calculation speedup of the TAAH tree algorithm relative to the FH tree algorithm.}
   \begin{tabular}{c|cccccc} 
    \hline
    \hline
     & $N_{L_2}$ & $N_{L_n}$ & $N_\mathrm{pixel}$ & $N_*$ & $\frac{t_\mathrm{SA}}{t_\mathrm{TAAH}}$ & $\frac{t_\mathrm{FH}}{t_\mathrm{TAAH}}$\\ 
    \hline
    Type I & $8.4 \times 10^5$  & $10 ^8$ & $1.3 \times 10^{10}$ & $8486$ & $1.4\times 10^4$ & $16$\\
    Type II & $1.4 \times 10^6$  & $1.6 \times 10^8$ & $2.3 \times 10^{10}$ & $14322$ & $1.5 \times 10^4$ & $10$\\
    Type III & $8.4 \times 10^5$  & $1.5 \times 10 ^8$ & $1.3 \times 10^{10}$ & $8486$ & $1.3\times 10^4$ & $15$  \\
    \hline
    \hline
  \end{tabular}
\end{table*}

\section{Population study}
\label{sec:population}
In this section, we utilize the TAAH code to explore the microlensing wave effect on SLGWs generated from stellar mass binary black holes (sBBHs). 
Here, we adopt the methodology outlined in~\citet{Shan:2023ngi,Shan:2023qvd} to simulate strong lensing and microlensing effects. 
Essentially, our approach assumes that the binary black hole merger rate is proportional to the star formation rate~\citep{Haris:2018vmn}. 
The criteria for strong lensing events are determined by the multi-image optical depth of the singular isothermal sphere (SIS) model. 
The microlensing density is aligned with the S\'{e}rsic light profile~\citep{Vernardos_2018}, the stellar mass function is derived from the Chabrier initial mass function~\citep{2003PASP..115..763C}, and the remnant mass function is based on the Spera initial-final relation~\citep{2015MNRAS.451.4086S}.
One can see Appendix~\ref{app:Strong_micro_simul} for more details.

\begin{figure*}
\centering
\includegraphics[width=0.8\textwidth]{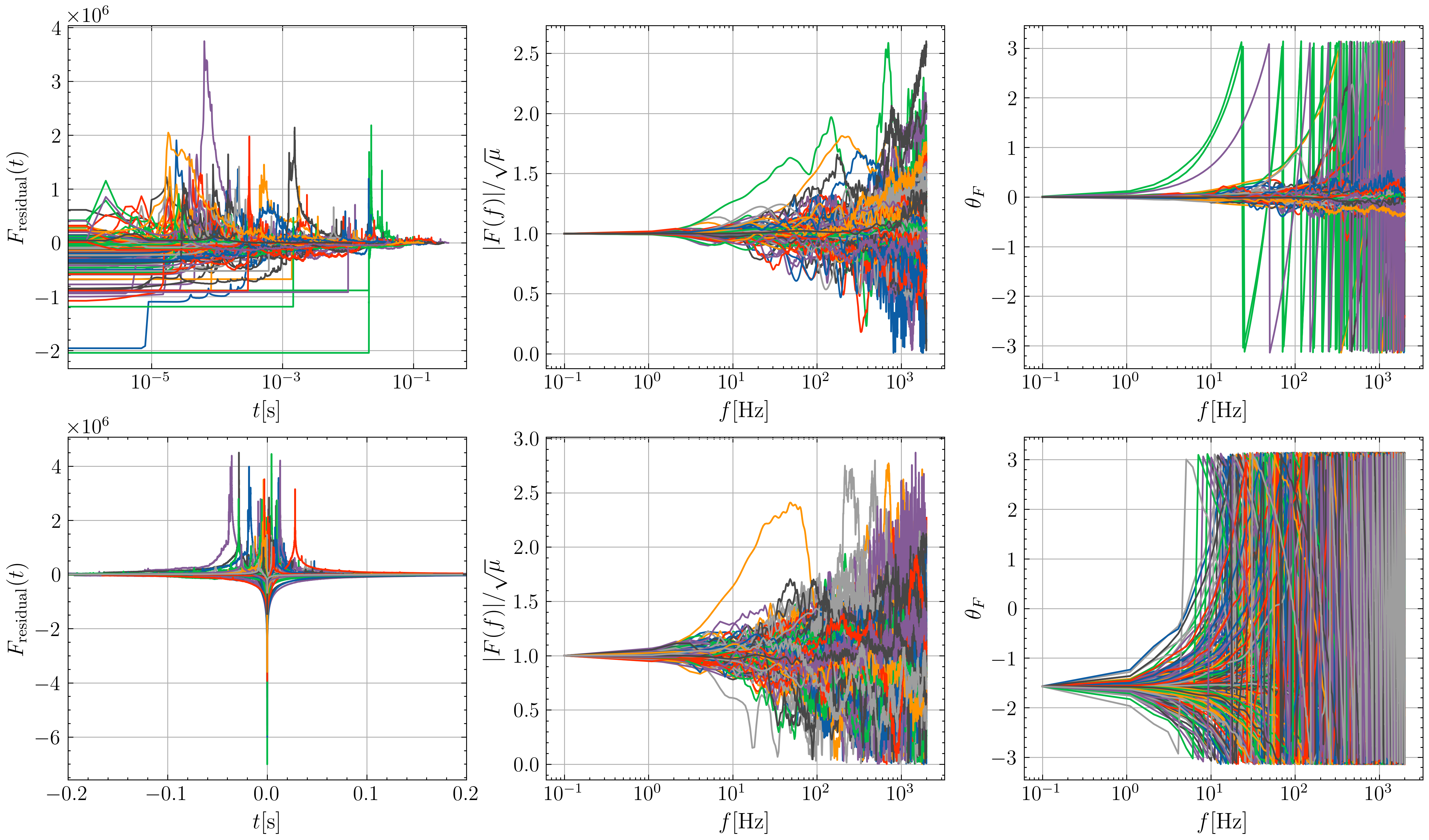}

\caption{This figure shows the microlensing diffraction results for SLGW events observed by 3 CE detectors over a span of two years observational time (approximately $250$ Type I and $250$ Type II events), different curves stand for different events. 
The first and second rows depict the results for Type I and Type II SLGWs, respectively. 
The first, second, and third columns display the residual time domain amplification factor ($F(t) - F_\mathrm{smooth}(t)$, where $F_\mathrm{smooth}(t)$ is the time domain amplification factor excluding microlenses), normalized amplification factor in the frequency domain, and complex phase, respectively. }
\label{fig:Ft_Ff_88_}
\end{figure*}

Fig.~\ref{fig:Ft_Ff_88_} illustrates the microlensing diffraction results for simulated SLGW events observed by three Cosmic Explorer (CE) detectors~\citep{Evans:2016mbw} (with network matched filter signal-to-noise ratio, SNR, $\geq 12$), situated at the locations of LIGO and Virgo, over a span of two years.
The data includes approximately $250$ Type I and $250$ Type II events, with a strong lensing event rate of approximately $10^{-3}$, which aligns with the findings of~\citet{Piorkowska:2013eww, Biesiada:2014kwa, Ding:2015uha, Li:2018prc, Yang:2019jhw, Yang:2021viz, Xu:2021bfn, Gao:2023uxi}, etc.
Different curves stand for different events.
The first and second rows depict the results for Type I and Type II SLGWs, respectively. 
The first, second, and third columns display the residual time domain amplification factor ($F(t) - F_\mathrm{smooth}(t)$, where $F_\mathrm{smooth}(t)$ is the time domain amplification factor excluding microlenses), normalized amplification factor in the frequency domain, and complex phase, respectively. 

From the first column, it is observed that the tail of each curve gradually approaches zero, indicating the convergence of the diffraction integral~\citep{Shan:2022xfx}. 
In the second and third columns, the wave effect in Type II SLGWs appears more pronounced than in Type I SLGWs, consistent with the findings of~\citep{2019Diego}. 
Moreover, fluctuations are noticeable in the frequency range of $10$ to $200$Hz, corresponding to the frequency range of GWs generated by sBBHs. 
Therefore, the anticipation of detecting microlensing fringes on ground-based detectors, particularly in third-generation ($3$G) detectors, seems reasonable.

\begin{figure}
\centering
\includegraphics[width=0.6\columnwidth]{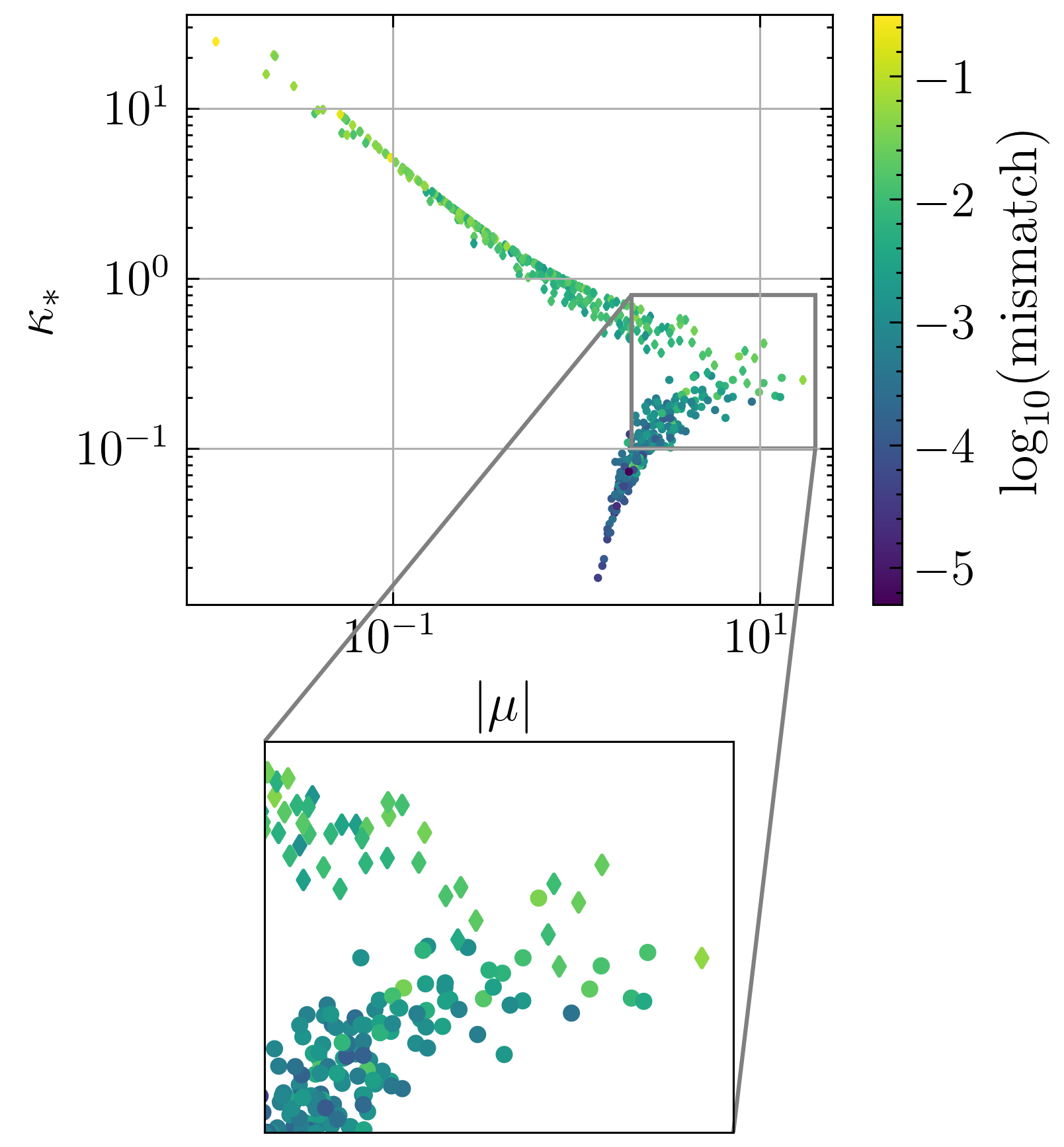}

\caption{This figure shows the mismatch analysis for all of the events shown in Fig.~\ref{fig:Ft_Ff_88_}.
The $x$-axis represents the absolute value of strong lensing magnification, while the $y$-axis represents the convergence or density of stellar microlensing field. 
The color bar indicates the logarithmic value of mismatch (defined in Eq.~\ref{eq:match}). 
Circle and diamond points represent Type I and Type II SLGWs, respectively. 
The zoomed-in panel shows the events with magnification larger than $2$.}
\label{fig:mismatch_88_}
\end{figure}

Fig.~\ref{fig:mismatch_88_} presents the mismatch analysis of these approximately $500$ SLGW events. 
The $x$-axis represents the absolute value of strong lensing magnification, while the $y$-axis represents the density of stellar microlensing field ($\kappa_*$). 
Circle and diamond points represent Type I and Type II SLGWs, respectively. 
The color bar indicates the logarithmic value of mismatch, defined as:
\begin{equation}
\label{eq:match}
\mathrm{mismatch} = 1 - \max_{\phi_0, t_0} \frac{\langle {h}_1 \mid {h}_2\rangle}{\sqrt{\langle {h}_1 \mid {h}_1\rangle\langle {h}_2 \mid {h}_2\rangle}} \
\end{equation}
where ${h}_1$ and ${h}_2$ are the waveform of signal $1$ and waveform of signal $2$, respectively.
$\phi_0$ and $t_0$ are the initial phase and start time of the signal $1$, respectively.
Therefore, this equation already takes into account the time delay from both macro and micro lensing.
$\langle . \mid . \rangle$ stands for the noise-weighted inner product and is defined as
\begin{equation}
\langle {h}_{1} \mid {h}_{2}\rangle=4 \operatorname{Re} \int_{f_{\text {low }}}^{f_{\text {high }}} \mathrm{d} f \frac{{h}_{1}^*(f)\times {h}_{2}(f)}{S_{\mathrm{n}}(f)} \;,
\end{equation}
where $S_\mathrm{n}(f)$ is the single-sided power spectral density of the detector noise, and ``$^*$'' denotes the complex conjugate value.
In this figure, signal $1$ is the unlensed waveform  ${h}_\mathrm{U}$ and signal $2$ is the macro + microlensing lensed waveform ${h}_\mathrm{L}$, respectively.

It is evident that strong lensing magnification and microlensing density both play crucial roles in the strength of the microlensing wave effect. In other words, higher magnification and density result in a more significant microlensing wave effect. This is because strong lensing magnification can increase the equivalent mass to $\mu M$~\citep{Diego_2018}, and a high-density field contains more high-mass microlenses. Therefore, SLGWs are more likely to interact with high-mass microlenses as they pass through the lens galaxy.

\begin{figure}
\centering
\includegraphics[width=0.6\columnwidth]{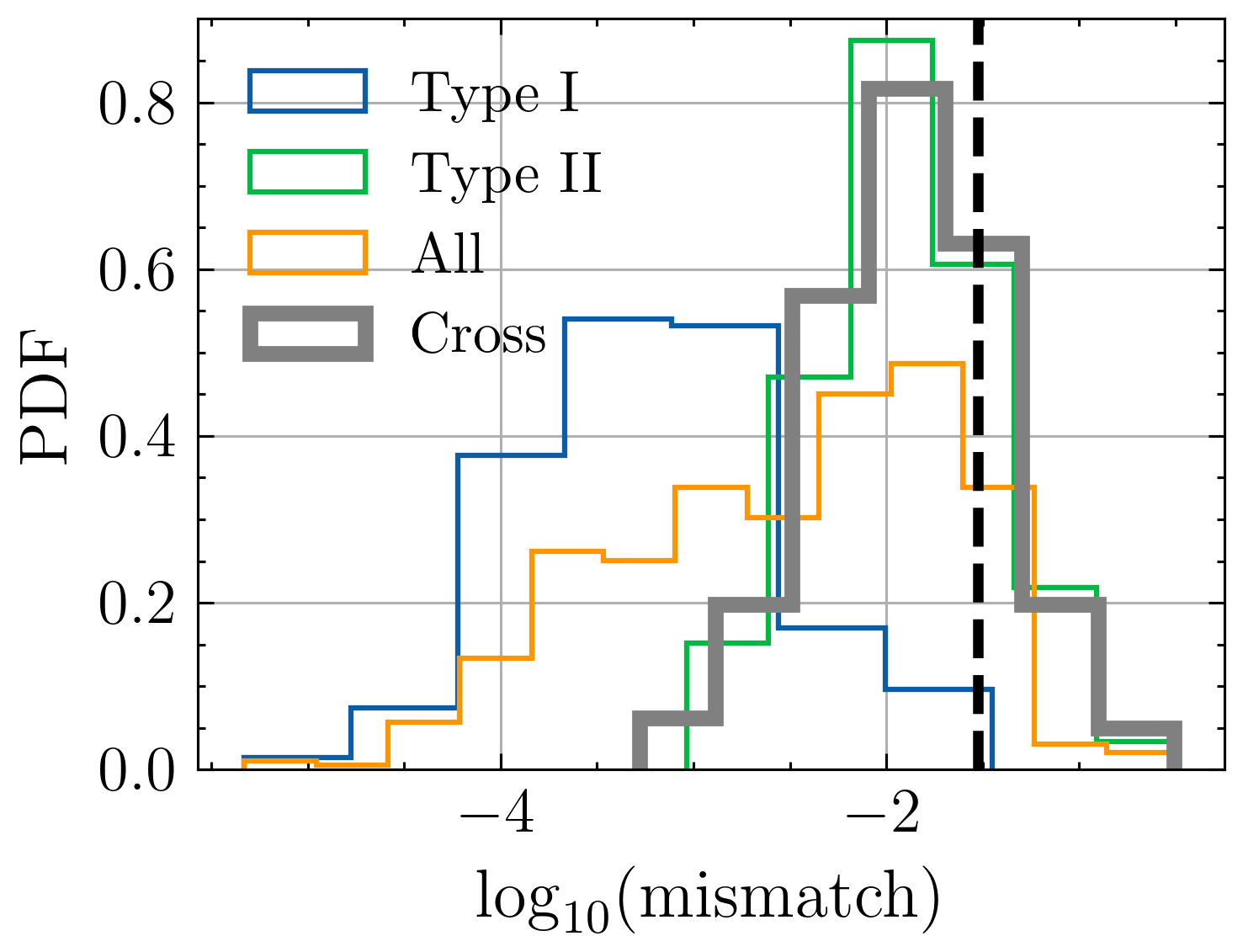}
\caption{This figure shows the mismatch histogram for all the events shown in Fig.~\ref{fig:Ft_Ff_88_}.
The blue, green, and orange curves represent the mismatch of Type I, Type II, and all events with respect to the unlensed waveform, respectively. 
The thick grey curve indicates the mismatch between signal pairs in a doubly imaged GW.
The vertical black dashed curve is the threshold, $3\%$, for template-bank spacing for GW detection~\citep{2005PhRvD..71f2001A, Harry_2011, Allen_2012}.}
\label{fig:pdf_mismatch_88_}
\end{figure}

Fig.~\ref{fig:pdf_mismatch_88_} shows the histogram of statistical results for the mismatch of events as presented in Fig.~\ref{fig:mismatch_88_}. 
The blue, green, and orange curves represent the mismatch of Type I, Type II, and all events with respect to the unlensed waveform, respectively. 
We find that more than $33\%$ of events have a mismatch larger than $1\%$, and $11\%$ of events have a mismatch larger than $3\%$ (vertical black dashed curve), which is a threshold for template-bank spacing for GW detection~\citep{2005PhRvD..71f2001A, Harry_2011, Allen_2012}. 
This means a mismatch larger than $3\%$ can lead to misidentification of GWs, not to mention parameter estimation bias.
Therefore, the microlensing wave optics effect is not negligible for SLGW, and one can expect to find microlensing evidence in the era of $3$G detectors (refer to \citealt{Shan:2023ngi} for a new detection method).

In addition, we calculated the mismatch between signal pairs in a doubly imaged GW (we did not include triply and quadruply imaged GW in this pair mismatch analysis step due to their relatively low event numbers compared to doubly imaged GW).
The results are shown by the thick grey curve. 
It can be seen that the mismatch between signal pairs aligns with the mismatch between Type II waveform and the unlensed waveform. 
Statistically, almost all strong lensing signal pairs have a mismatch larger than $10^{-3}$, with $61\%$ of signal pairs having a mismatch larger than $1\%$ and $25\%$ having a mismatch larger than $3\%$.
Therefore, the microlensing-induced mismatch can reduce the power of SLGW identification based on an overlapping method, especially for more sensitive detecors (refer to \citealt{Shan:2023qvd} for a similar conclusion).

In this figure, it is also clear that the mismatch of Type II SLGWs is higher than that of Type I SLGWs. However, this figure does not indicate whether the larger mismatch of Type II SLGWs is due to their distinctive geometric structure or because Type II images are usually situated in regions of higher density. As shown in Fig.~\ref{fig:mismatch_88_}, the microlensing density of Type II events is generally higher than that of Type I events. The zoomed-in panel in Fig.~\ref{fig:mismatch_88_} provides some weak evidence that the larger mismatch of Type II SLGWs is due to their intrinsic geometric properties, as indicated by the comparison of two closely situated circle and diamond points.

To confirm this conclusion, we selected one Type I and one Type II configuration with similar strong lensing magnification and microlensing density from the population simulation step. The convergence and shear of each Type I SLGW are approximately $0.596$, with a microlensing density $\kappa_* \simeq 0.28$. The convergence and shear of each Type II SLGW are approximately $0.403$, with a microlensing density $\kappa_* \simeq 0.28$. Under this configuration, the strong lensing magnifications of both Type I and Type II SLGWs are close to $5.2$. 
We then simulated $50$ new Type I SLGWs and $50$ new Type II SLGWs under these two configurations, respectively.

\begin{figure*}
\centering
\includegraphics[width=0.8\textwidth]{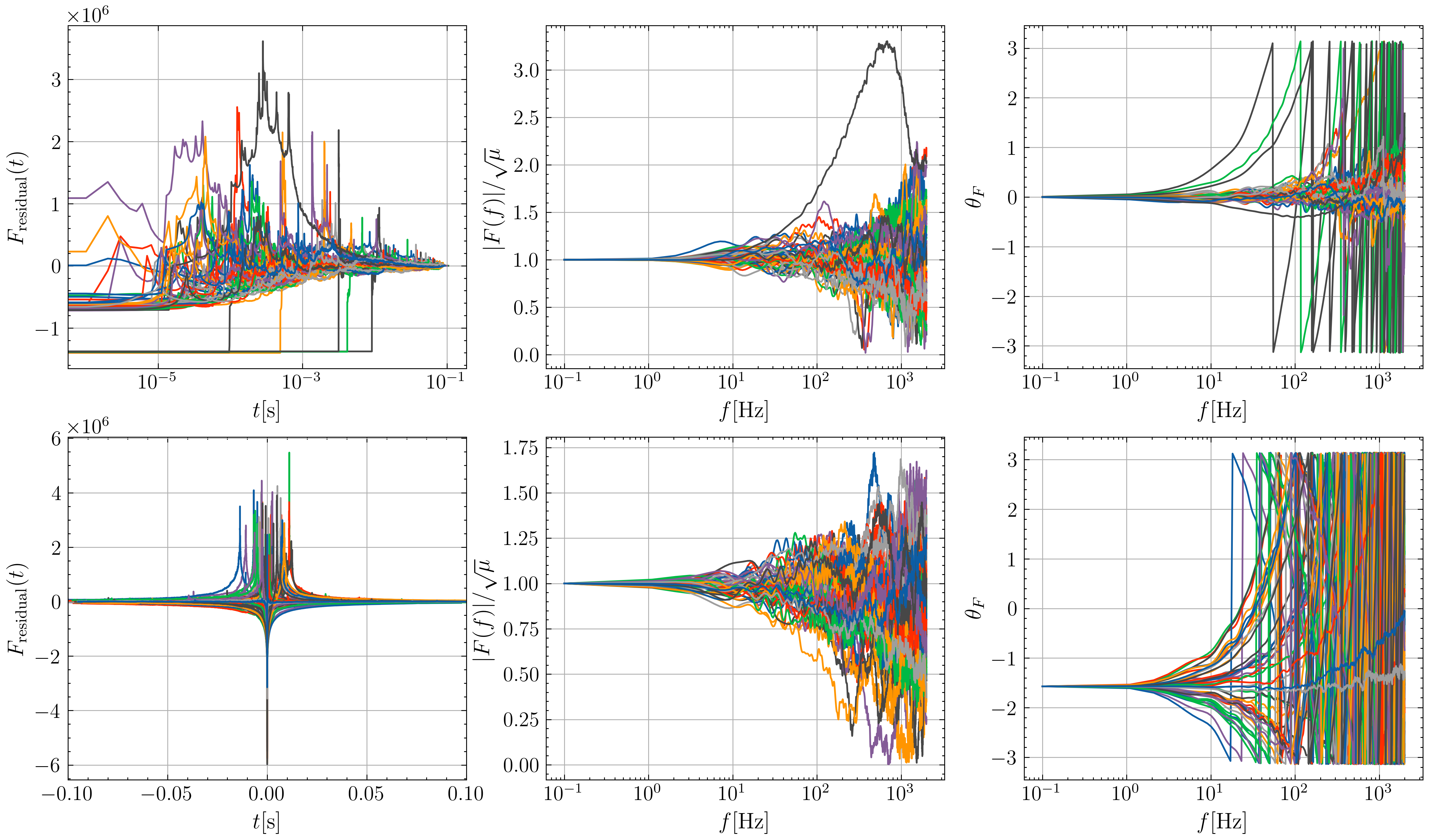}

\caption{This figure shows the diffraction integral results for $50$ Type I and $50$ Type II SLGW events for mismatch comparision. The first and second rows depict the results for Type I and Type II SLGWs, respectively, with different curves representing different events. The first, second, and third columns display the residual time domain amplification factor, normalized amplification factor in the frequency domain, and complex phase, respectively. }
\label{fig:Ft_Ff_50_}
\end{figure*}

Fig.~\ref{fig:Ft_Ff_50_} shows the diffraction integral results for these $100$ SLGW events. The first and second rows depict the results for Type I and Type II SLGWs, respectively, with different curves representing different events. The first, second, and third columns display the residual time domain amplification factor, normalized amplification factor in the frequency domain, and complex phase, respectively.

\begin{figure}
\centering
\includegraphics[width=0.6\columnwidth]{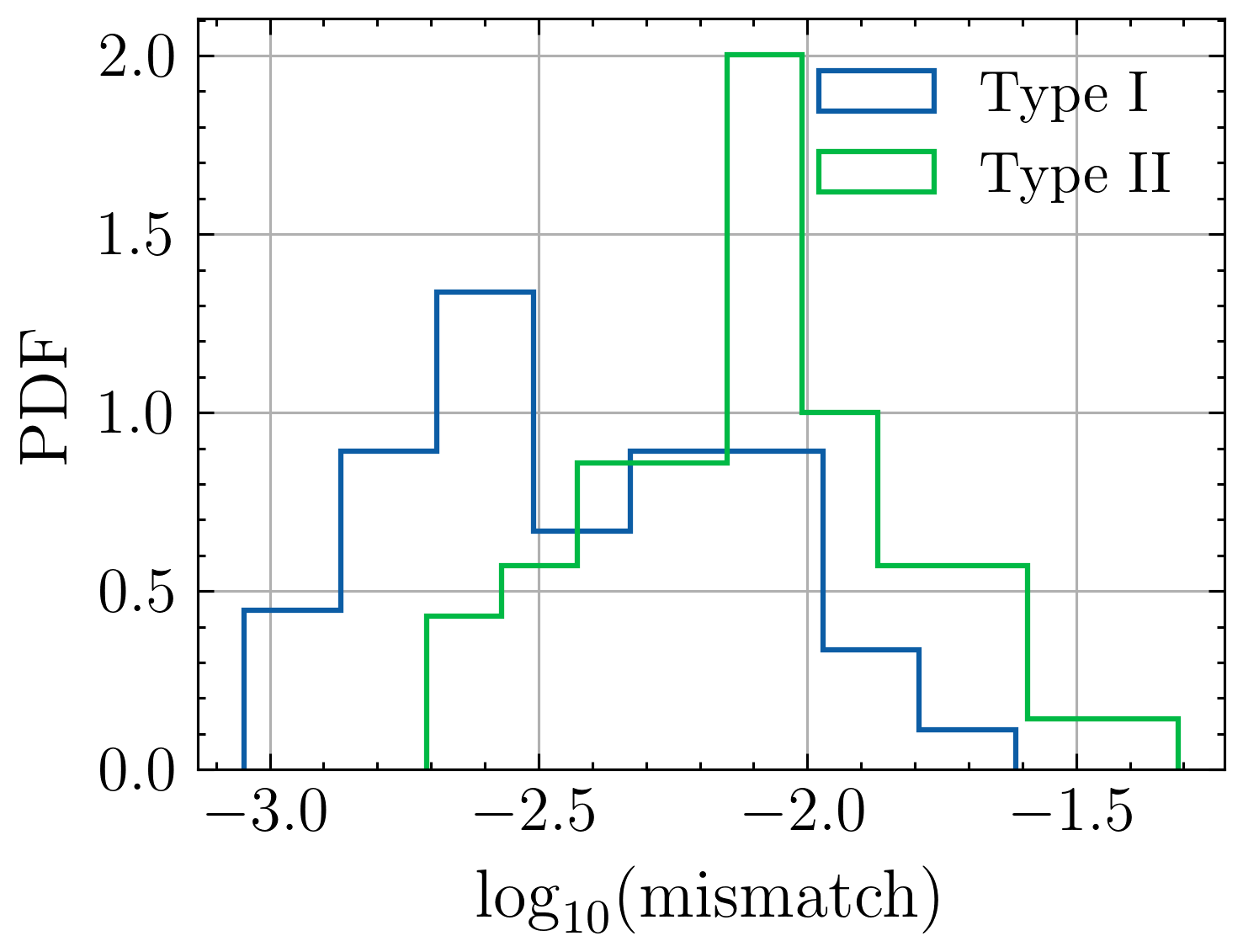}

\caption{This figure shows the mismatch probability distribution of events shown in Fig.~\ref{fig:Ft_Ff_50_}.
The blue and green curves represent the Type I and Type II events, respectively. }
\label{fig:mismatch_50_}
\end{figure}

Fig.~\ref{fig:mismatch_50_} presents the mismatch probability distribution function of these $100$ events. The blue and green curves represent the mismatch between the Type I waveform and the unlensed waveform, and the Type II waveform and the unlensed waveform, respectively. It is evident that the mismatch for Type II events is significantly larger than that for Type I events, under the similar magnification and microlensing configuration. Therefore, the significant mismatch of Type II events compared with Type I events in Fig.~\ref{fig:mismatch_88_} is not only due to the higher microlensing density but also their unique geometric properties.

\section{conclusion and discussion}
\label{sec:condis}
The gravitational lensing wave effect induced by the microlensing field within the lens galaxy presents both challenges and opportunities for the detection and application of SLGW. 
Although various algorithms have been proposed to address the oscillation diffraction integral~\citep{UG95, press1992numerical, levin1982procedures, filon1930iii, xiang2007efficient, iserles2006computation, guo2020convergence, Tambalo:2022plm}, applying them directly to the microlensing field scenario remains challenging. 
This difficulty arises from the multitude of microlenses in this field, numbering in the thousands or even millions, which leads to numerical calculations inevitably encountering issues with convergence, precision, and time consumption.

As a continuation of our prior work~\citep{Shan:2022xfx}, where we tackled convergence problems and enhanced numerical precision using a component decomposition algorithm, this paper introduces a Trapezoid Approximation-based Adaptive Hierarchical (TAAH) tree algorithm within the pixel. 
This algorithm effectively addresses the challenges related to lens plane resolution and computational time.

In Fig.~\ref{fig:OneMicroFf_ThetaF}, we compare our TAAH tree algorithm with the theoretical formula and traditional Simple Adding (SA) algorithm under a microlens embedded in Type I (minimum), Type III (maximum), and pure Type II (saddle) scenarios. 
Our findings reveal that the TAAH tree algorithm achieves comparable or higher accuracy than the traditional SA algorithm while avoiding the need for high density sampling in the lens plane. 
More importantly, the TAAH tree algorithm provides a numerical error control method, resulting in more robust results.

Subsequently, we apply our TAAH tree algorithm to the microlensing field scenario. 
In Fig.~\ref{fig:MicroFieldMesh}, we illustrate the resolution levels of the TAAH tree algorithm. 
We observe that only a small fraction of pixels require refinement relative to the Level $2$ grid, resulting in faster calculations compared to the Fixed Hierarchical (FH) tree algorithm. 
As detailed in Table~\ref{ta:MultiResult}, the TAAH tree algorithm exhibits a four orders of magnitude acceleration compared to the SA algorithm and one order of magnitude acceleration compared to the FH tree algorithm. 
In Fig.~\ref{fig:MicroFieldFf_ThetaF}, we present the results obtained with the TAAH tree algorithm and the numerical differences between TAAH tree and FH tree algorithms. 
Our algorithm demonstrates a maximum difference of only $1\%$.

Finally, we utilized our TAAH code to explore the microlensing wave effect on SLGWs generated from sBBHs. 
This analysis was conducted on a more comprehensive sample set of approximately $250$ Type I SLGWs and $250$ Type II SLGWs. 
Our investigation yielded four findings:
\begin{itemize}[label=\textbullet, font=\Large]
    \item The strength of the microlensing wave effect depends on both the strong lensing magnification and the microlensing field density.
    \item The microlensing effect on Type II SLGWs, with a potential mismatch exceeding $10\%$, is greater than that on Type I SLGWs, where the mismatch is typically ($95\%$ events) below $1\%$. The higher mismatch of Type II events is not only due to their interaction with a denser microlensing field but also because of the unique intrinsic geometric structure of the Type II signal, based on the results in Fig.~\ref{fig:Ft_Ff_50_} and Fig.~\ref{fig:mismatch_50_}.
    \item More than $33\%$ of events have a mismatch larger than $1\%$, and $11\%$ of events have a mismatch larger than $3\%$.
    \item The mismatch between signal pairs in a doubly imaged GW is generally larger than $10^{-3}$. Additionally, $61\%$ of signal pairs have a mismatch larger than $1\%$, and $25\%$ have a mismatch larger than $3\%$. Therefore, the microlensing-induced mismatch can reduce the SLGW identification ability using the overlapping method.
\end{itemize}
Based on these findings, it is evident that the microlensing wave effect cannot be disregarded, particularly for Type II SLGWs, and one can expect to find microlensing evidence in the future, such as in the era of $3$G detectors.

This finding differs from the results in \citet{2021Anuj, Meena:2022unp}, which stated $>99\%$ events have a mismatch less than $1\%$, due to three main reasons:
\begin{enumerate}[label=\textbullet, font=\Large]
    \item Our study includes a more comprehensive sample set, considering all SLGWs rather than focusing solely on moderate density and moderate strong lensing magnification. This broader inclusion provides a more holistic view of the microlensing wave effect.
    \item Unlike \citet{Meena:2022unp}, who used a fixed microlensing field boundary of $2~\mathrm{pc} \times 2~\mathrm{pc}$ for all SLGWs, we selected the boundary by evaluating the convergence of the diffraction integral. Our approach revealed that the convergence number of microlenses is often $>10^5$ under a diffraction integral SNR configuration of $100$~\citep{Shan:2022xfx}, significantly higher than the maximum number of around $10^4$ used in \citet{Meena:2022unp}.
    \item We did not exclude microlenses with masses less than $0.2~\mathrm{M_\odot}$ outside a $1.5~\mathrm{pc}^2$ area, as was done in \citet{2021Anuj}.
\end{enumerate}



In summary, we have introduced a novel numerical algorithm for the diffraction integral, specifically tailored for the microlensing field within the lens galaxy. 
This algorithm effectively resolves numerical resolution challenges and provides control over numerical errors. 
Furthermore, the incorporation of an adaptive hierarchical tree algorithm enhances computational efficiency.
Finally, we applied this algorithm to analyze the microlensing wave effect from a statistical point of view. 
We found that the microlensing wave effect cannot be ignored, especially for Type II SLGWs, and one can expect to identify the microlensing signature in the era of $3$G detectors.

\vspace*{2mm} \Acknowledgements{\bahao  }
This work is supported by the NSFC (No. U1931210, 11673065, 11273061). We acknowledge the science research grants from the China Manned Space Project with NO.CMS-CSST-2021-A11.
B.H. acknowledges support by the National Natural Science Foundation of China (grant No. 12333001).
S.M. acknowledges support by the National Natural Science Foundation of China (grant No. 12133005).
%




\appendix
\section{Strong lensing and microlensing population simulation}
\label{app:Strong_micro_simul}
This appendix details the simulation procedure in Section~\ref{sec:population}.
To validate the method, we follow the procedures outlined in~\citet{Haris:2018vmn,Xu:2021bfn,Shan:2023qvd} to generate a mock dataset using the Monte Carlo method. 
The primary simulation process unfolds as follows:

\begin{itemize}[label=\textbullet, font=\Large]
  \item
  We sample the sBBH redshift from a theoretical sBBH merger model in which the merger rate is proportional to the star formation rate with a delay time $\Delta t = 50 \mathrm{Myr}$ between the star and sBBH formation. The details can be found in~\citet{Shan:2023ngi,Shan:2023qvd}.
  \item
  For the events picked above, we randomly assign sBBH masses ($m_1$, $m_2$), inclination angle ($\iota$), polarization angle ($\psi$), right ascension angle ($\alpha$), declination ($\delta$), merger time ($t_c$), and spins ($a_1$, $a_2$) from the following distributions. 
      \begin{itemize}
      \item [a)]
      $(m_1, m_2)\sim \mathrm{power\ law + peak}$~\citep{LIGOScientific:2018jsj}.
      \item [b)]
      $p(\iota)\propto \sin(\iota)$, $\iota \in  [0, \pi]$.
      \item [c)]
      $p(\psi)\propto \mathrm{U}(0,\pi)$.
      \item [d)]
      $p(\alpha)\propto \mathrm{U}(0,2\pi)$.
      \item [e)]
      $p(\delta)\propto \cos(\delta)$, $\delta \in [-\pi/2, \pi/2]$.
      \item [f)]
      $p(t_c)\propto \mathrm{U}(t_\mathrm{min}, t_\mathrm{max})$, where $t_\mathrm{min}$ and $t_\mathrm{max}$ are the minimum and maximum merger times used in the simulation. Here, we set $t_\mathrm{max}-t_\mathrm{min}=2\mathrm{yr}$.
      \item [g)]
      $p(a_1)\propto \mathrm{U}(0, 0.99)$.
      \item [h)]
      $p(a_2)\propto \mathrm{U}(0, 0.99)$.
      \end{itemize}
  \item
  Calculate the multiple-image optical depth $\tau(z_s)$ for each sBBH at redshift $z_s$ using the SIS optical depth as shown in~\citet{Haris:2018vmn,Shan:2023qvd}. 
  Then, generate a random number uniformly distributed between $0$ and $1$ for each sBBH event. 
  Compare the calculated optical depth $\tau(z_s)$ with the generated random number for each event. 
  If the optical depth $\tau(z_s)$ is larger than the random number, classify it as an SLGW event; otherwise, exclude it from the selection. 
   \item
  For the selected SLGW samples, we assume a singular isothermal ellipsoid (SIE) lens model~\citep{1994A&A...284..285K} and use \texttt{Lenstronomy}~\citep{2018PDU....22..189B, 2021JOSS....6.3283B} to solve the lens equation.
  The velocity dispersion $\sigma_v$ and axis ratio $q$ of SIE are generated from the SDSS galaxy population distribution~\citep{2015ApJ...811...20C,Wierda:2021upe}.
   \item
  Then, we introduce the detector's selection effect in the above samples.
  We use three CE detectors, located at Livingston, Hanford, and Pisa to receive GW signals and assume that an event can be detected if its network matched filter SNR $\geq 12$.
  By using these three detectors, one can potentially observe approximately $\sim 3\times10^{5}$ sBBHs and $250$ strong lensing systems in $2$ years observational time. 
  This result aligns with the findings of~\citet{Piorkowska:2013eww, Biesiada:2014kwa, Ding:2015uha, Li:2018prc, Yang:2019jhw, Yang:2021viz, Xu:2021bfn, Gao:2023uxi}, etc, which reported that the rate of strong lensing events approximately ranges from $10^{-3}$ to $10^{-4}$.
   \item
  Finally, our focus shifts to the simulation of the microlensing field. 
  In this study, we utilize the Chabrier initial mass function~\citep{2003PASP..115..763C} and an elliptical S\'{e}rsic profile~\citep{Vernardos_2018} to describe the stellar mass function and density associated with each SLGW. 
  Specifically, we set the stellar mass range to be within $[0.1, 1.5]$ solar masses, which aligns with the value employed in~\citet{Diego:2021mhf}. 
  In addition to the stellar mass component, we also consider the presence of remnant objects in the microlensing field. 
  We adopt the initial-final relation from~\citet{2015MNRAS.451.4086S} to describe the remnant mass function. 
  The remnant mass density has been set at $10\%$ of the stellar mass density~\citep{Meena:2022unp}.
\end{itemize}



\bibliographystyle{unsrtnat}
\bibliography{ref}





\end{document}